\documentstyle[aps]{revtex}

\begin{document}
\title{Disentangling and broadcasting an entangled state simultaneously by
asymmetric cloning}
\author{Yafei Yu$^{\thanks{%
corresponding author, e-mail: yfyu@wipm.ac.cn}a}$, Jian Feng$^{a,b}$,
Xiaoqing Zhou$^{a}$, Mingsheng Zhan$^{a}$}
\address{$^{a}$State Key Laboratory of Magnetic Resonance and Atomic and Molecular\\
Physics,\\
Wuhan Institute of Physics and Mathematics, Chinese Academy of Sciences, \\
$^{b}$Institute of Optical Communication, Liaocheng University,Liaocheng ,\\
252059, Shandong , PR China}
\maketitle

\begin{abstract}
We construct a quantum machine which, by using asymmetric cloner, deals with
disentangling and broadcasting entanglement in a single unitary evolution.
The attainable maximum value of the scaling parameter $s$ for disentangling
is identical to that obtained in previous works. The fidelity of the cloning
state with respect to the input entangled state is state-dependent.

PACS number(s): 03.67.HK
\end{abstract}

\section{ Introduction}

Quantum entanglement plays an important role in quantum information field.
And the manipulation of entanglement, such as purification \cite{bch},
broadcasting \cite{87,85}, is an intriguing issue in the research for
entanglement. Recently disentanglement attracts a lot of attention.
Disentanglement is a process in which an initial entangled state of
composite system can be transformed into separable state with unaffecting
the reduced density matrices of subsystems. However like other ''no-going''
theorems ( e.g., the no-cloning theorem\cite{1}, no-deleting theorem\cite{2}%
, no-broadcasting theorem \cite{3}), the perfect disentanglement also is
prohibited by elementary rules of quantum mechanics \cite{83,82}. But
dropping the constraint that the reduced density matrices of subsystems is
perfectly unaffected, the approximate disentanglement can be realized by
local operation \cite{88}, for example, by local cloning\cite{7} and by
teleportation via separable channels\cite{8}. In other words, for a
two-qubit entangled state, the transformation can be achieved:

\begin{equation}
\rho ^{ent}\rightarrow \rho ^{disent}
\end{equation}

together with

\begin{equation}
Tr_{j}(\rho ^{disent})=s_{i}Tr_{j}(\rho ^{ent})+(\frac{1-s_{i}}{2})I,i\neq
j;i,j=1,2
\end{equation}
for all $\rho ^{ent}$, where $s_{i}$ ($0<s_{i}<1$ for $i=1,2$) is a scaling
parameter independent of $\rho ^{ent}$, standing as a measure of closeness
of the $ith$ reduced density matrix before and after the transformation. The
attainable values of $s_{1}$ and $s_{2}$ satisfies the inequality $%
s_{1}s_{2}\leq \frac{1}{3}$. More explicitly, if only one party undergoes
local operation, i.e., $s_{1}=1$ (or $s_{2}=1$), the maximum value of $s_{2}$
(or $s_{1}$) is $\frac{1}{3}$; if both two parties undergo the same local
operation separately, the maximum value of $s_{2}$ (or $s_{1}$) is $\frac{1}{%
\sqrt{3}}$.

On the other hand, the connection between copying (cloning and broadcasting)
and disentanglement is noted\cite{82}. It is found that the conditions of
perfect disentanglement into product states corresponds to cloning of one of
the subsystems while the conditions of perfect disentanglement into
separable states corresponds to broadcasting of one of the subsystems\cite
{84}. And comparing the schemes of disentanglement and broadcasting
entanglement by local symmetric cloning\cite{85,86}, one can notice that for
disentanglement the reduction factor $\eta $ describing the quality of
symmetric cloner satisfies $\eta \leq \frac{1}{\sqrt{3}}$, whereas for
broadcasting it needs $\eta \geq \frac{1}{\sqrt{3}}$ ($\eta =\frac{2}{3}$
for the optimal quantum cloner\cite{83(3)}). In the meantime,
disentanglement erases the quantum correlation (inseparability) between
subsystems as broadcasting needs to retain the quantum correlation between
subsystems, except that in both process the output reduced density matrix of
each subsystem is as close as possible to the corresponding input reduced
density matrix. There is an intuitive understanding that the disentangling
state of an entangled state can be viewed as a separable copy of it. This
suggests that it is possible to look for a unified way of dealing with
broadcasting and disentangling an entangled state simultaneously.

In this short note, we substitute asymmetric (isotropic) cloner for the
symmetric (isotropic) cloner in the schemes of broadcasting entanglement and
disentanglement by local cloning \cite{85,86}. Therefore a quantum machine
is constructed, which can implement the broadcasting and disentanglement of
an entangled state in a single unitary evolution. In Section II we first
discuss the mechanics of the quantum machine in the method similar to
entanglement splitting \cite{99}, then consider its functions in the
framework of entanglement broadcasting \cite{87,85}. Finally, the
conclusions is given in Section III.

\section{Disentangling and broadcasting by asymmetric cloning}

In this section we will analyze the actions of the quantum machine which
deals with disentangling and broadcasting an entangled state simultaneously.
The mechanics of the quantum machine is discussed in the method similar to
entanglement splitting. Then the functions of the quantum is considered in
the framework of entanglement broadcasting. That is, the discussion is
divided into two cases, using the cloner on only one party and on both
parties separately. For simplicity, the discussion focuses on the case of
qubit, and only 1$\rightarrow $2 asymmetric (isotropic) cloner is exploited.

The actions of asymmetric cloner in general is described by a Pauli channel%
\cite{acta,nj}. But, for convenience, here the actions of asymmetric cloner
is specified by a particular unitary transformation on the state $\left|
\varphi \right\rangle =\alpha \left| 0\right\rangle +\beta \left|
1\right\rangle $ of the input qubit $a$, where $\alpha $ and $\beta $ are
unknown real parameters, $\alpha ^{2}+\beta ^{2}=1$ and $0\leq \alpha ,\beta
\leq 1$. From orthogonality, unitarity and isotropy, this transformation can
be presented as: 
\begin{equation}
\left| 0\right\rangle _{a}\left| \xi \right\rangle _{b}\left| Q\right\rangle
_{c}\rightarrow \frac{1}{\sqrt{N}}\left| 00\right\rangle _{ab}\left|
\uparrow \right\rangle _{c}+(\frac{p}{\sqrt{N}}\left| 01\right\rangle _{ab}+%
\frac{q}{\sqrt{N}}\left| 10\right\rangle _{ab})\left| \downarrow
\right\rangle _{c}
\end{equation}
\begin{equation}
\left| 1\right\rangle _{a}\left| \xi \right\rangle _{b}\left| Q\right\rangle
_{c}\rightarrow \frac{1}{\sqrt{N}}\left| 11\right\rangle _{ab}\left|
\downarrow \right\rangle _{c}+(\frac{p}{\sqrt{N}}\left| 10\right\rangle
_{ab}+\frac{q}{\sqrt{N}}\left| 01\right\rangle _{ab})\left| \uparrow
\right\rangle _{c},
\end{equation}
where $N$ is a normalization factor given by $N=1+p^{2}+q^{2}$, $q=1-p,$ $%
p>q $. $\left| Q\right\rangle _{c}$ describes the initial state of the
cloner, $\left| \xi \right\rangle _{b}$ is an arbitrary initial state of
qubit $b$, and $\left| \uparrow \right\rangle _{c}$ and $\left| \downarrow
\right\rangle _{c}$ are two orthonormal vectors in the Hilbert space of the
cloner. The asymmetric cloning transformation outputs two copies of the
state $\left| \varphi \right\rangle $ on the qubits$\ a$ and $b,$ 
\begin{equation}
\rho _{a}=\frac{2p}{N}\left| \varphi \right\rangle \left\langle \varphi
\right| +(1-\frac{2p}{N})\frac{I}{2}\text{,}
\end{equation}
\begin{equation}
\rho _{b}=\frac{2q}{N}\left| \varphi \right\rangle \left\langle \varphi
\right| +(1-\frac{2q}{N})\frac{I}{2}\text{,}
\end{equation}
where $I$ is unitary matrix. So for the qubits $a$ and $b$, the reduction
factors by which the cloner shrinks the vector characterizing the input
state in the Bloch sphere are $\eta _{a}=\frac{2p}{N}$ and $\eta _{b}=\frac{%
2q}{N}$, respectively. Now that $p\neq q$, the asymmetric cloner outputs two
copies with different fidelities for all input states. And 
\begin{equation}
\eta _{a}^{2}+\eta _{b}^{2}+\eta _{a}\eta _{b}-\eta _{a}-\eta _{b}=0,
\end{equation}
which satisfies the no-cloning inequality deduced from Eq. [6] in \cite{nj}.
Therefore the distribution of information at the outputs of the cloner is
controlled via changing the value of $p$\cite{177}.

In the following this family of asymmetric cloners will be used to obtain a
disentangling state and a broadcasting state of an entangled state
simultaneously.

\subsection{only copying one qubit asymmetrically}

We first discuss why the asymmetric cloner can be used to disentangle and
broadcast entanglement. The discussion is in the method similar to
entanglement splitting, i.e., only one qubit is copied by asymmetric cloning.

Suppose qubits $a_{I}$ and $a_{II}$ share an entangled state

\begin{equation}
\begin{array}{c}
\left| \chi \right\rangle =\alpha \left| 00\right\rangle
_{a_{I}a_{II}}+\beta \left| 11\right\rangle _{a_{I}a_{II}},
\end{array}
\end{equation}
where $\alpha $ and $\beta $ are defined as before. Let the asymmetric
cloner copy one qubit (say, qubit $a_{II}$) as doing in entanglement
splitting and output two copies on the qubits $a_{II}$ and $b_{II}$. Then
the original state $\left| \chi \right\rangle $ is splitted into two
branches. To the purpose of dealing with disentanglement and entanglement
broadcasting simultaneously, the quantum correlation in one branch is erased
whereas that in another needs to be partially retained. The state of each
branch can be described as: 
\begin{equation}
\begin{array}{c}
\rho _{a_{I}a_{II}}^{out}=\frac{(1+p^{2})}{N}\alpha ^{2}\left|
00\right\rangle \left\langle 00\right| +\frac{(1+p^{2})}{N}\beta ^{2}\left|
11\right\rangle \left\langle 11\right| +\frac{q^{2}}{N}\alpha ^{2}\left|
01\right\rangle \left\langle 01\right| \\ 
+\frac{q^{2}}{N}\beta ^{2}\left| 10\right\rangle \left\langle 10\right| +2%
\frac{p}{N}\alpha \beta \left| 00\right\rangle \left\langle 11\right| +2%
\frac{p}{N}\alpha \beta \left| 11\right\rangle \left\langle 00\right|
\end{array}
\end{equation}
\begin{equation}
\begin{array}{c}
\rho _{a_{I}b_{II}}^{out}=\frac{(1+q^{2})}{N}\alpha ^{2}\left|
00\right\rangle \left\langle 00\right| +\frac{(1+q^{2})}{N}\beta ^{2}\left|
11\right\rangle \left\langle 11\right| +\frac{p^{2}}{N}\alpha ^{2}\left|
01\right\rangle \left\langle 01\right| \\ 
+\frac{p^{2}}{N}\beta ^{2}\left| 10\right\rangle \left\langle 10\right| +2%
\frac{q}{N}\alpha \beta \left| 00\right\rangle \left\langle 11\right| +2%
\frac{q}{N}\alpha \beta \left| 11\right\rangle \left\langle 00\right| .
\end{array}
\end{equation}
By applying the Peres-Horodeck theorem\cite{12,13} to test the
inseparability of $\rho _{a_{I}a_{II}}^{out}$ and $\rho _{a_{I}b_{II}}^{out}$%
, it turns out that if we require that $\rho _{a_{I}a_{II}}^{out}$ is
inseparable while $\rho _{a_{I}b_{II}}^{out}$ is separable for any input
pure two-qubit entangled state, the values of $p$ must satisfy $\sqrt{3}%
-1\leq p\leq 1$. That is, $\rho _{a_{I}a_{II}}^{out}$ is a copying state of
the initial entangled state $\left| \chi \right\rangle $ while $\rho
_{a_{I}b_{II}}^{out}$ is a disentangling state of it when the parameter $p$
specifying the asymmetric cloning takes value between $\sqrt{3}-1$ and$\ 1$.

When $\alpha =\beta =\frac{1}{\sqrt{2}}$, two output density matrices can be
rewritten as: 
\begin{equation}
\begin{array}{c}
\rho _{a_{I}a_{II}}^{out}=\frac{1}{2}[\frac{(1+p^{2})}{N}\left|
00\right\rangle \left\langle 00\right| +\frac{(1+p^{2})}{N}\left|
11\right\rangle \left\langle 11\right| +\frac{q^{2}}{N}\left|
01\right\rangle \left\langle 01\right| \\ 
+\frac{q^{2}}{N}\left| 10\right\rangle \left\langle 10\right| +2\frac{p}{N}%
\left| 00\right\rangle \left\langle 11\right| +2\frac{p}{N}\left|
11\right\rangle \left\langle 00\right| ] \\ 
=\frac{(1+p)^{2}}{2N}\left| \Phi ^{+}\right\rangle \left\langle \Phi
^{+}\right| +\frac{(1-p)^{2}}{2N}\left| \Phi ^{-}\right\rangle \left\langle
\Phi ^{-}\right| \\ 
+\frac{(1-p)^{2}}{2N}\left| \Psi ^{+}\right\rangle \left\langle \Psi
^{+}\right| +\frac{(1-p)^{2}}{2N}\left| \Psi ^{-}\right\rangle \left\langle
\Psi ^{-}\right| ,
\end{array}
\end{equation}
\begin{equation}
\begin{array}{c}
\rho _{a_{I}b_{II}}^{out}=\frac{1}{2}[\frac{(1+q^{2})}{N}\left|
00\right\rangle \left\langle 00\right| +\frac{(1+q^{2})}{N}\left|
11\right\rangle \left\langle 11\right| +\frac{p^{2}}{N}\left|
01\right\rangle \left\langle 01\right| \\ 
+\frac{p^{2}}{N}\left| 10\right\rangle \left\langle 10\right| +2\frac{q}{N}%
\left| 00\right\rangle \left\langle 11\right| +2\frac{q}{N}\left|
11\right\rangle \left\langle 00\right| ] \\ 
=\frac{(2-p)^{2}}{2N}\left| \Phi ^{+}\right\rangle \left\langle \Phi
^{+}\right| +\frac{p^{2}}{2N}\left| \Phi ^{-}\right\rangle \left\langle \Phi
^{-}\right| \\ 
+\frac{p^{2}}{2N}\left| \Psi ^{+}\right\rangle \left\langle \Psi ^{+}\right|
+\frac{p^{2}}{2N}\left| \Psi ^{-}\right\rangle \left\langle \Psi ^{-}\right|
.
\end{array}
\end{equation}
According to the viewpoint in \cite{acta,nj}, it is clear that the qubits $%
a_{II}$ and $b_{II}$ emerge from depolarizing channels of probability $P=3%
\frac{(1-p)^{2}}{2N}$ and $P^{^{\prime }}=3\frac{p^{2}}{2N}$, respectively.
Hence the variation of the parameter $p$ changes the capacities of two
quantum channels such that the quantum correlation of the initial entangled
state $\left| \chi \right\rangle $ is filtered in the branch $a_{I}b_{II}$
but partially transferred to the branch $a_{I}a_{II}$. And it is noted that
the cloner of $\sqrt{3}-1\leq p\leq 1$ corresponds to the case of $%
x^{^{\prime }}\geq \frac{1}{\sqrt{6}}$ in Fig.2 in Ref. \cite{nj}.

Hence the flow of information in the asymmetric cloning is controlled by
varying the value of $p$ of the cloner. Because of this ability,
disentangling and broadcasting an entangled state is possible to be achieved
in a single unitary evolution.

\subsection{copying both qubits separately}

In the previous subsection we have discussed the mechanics of the quantum
machine which does with disentangling and broadcasting an entangled state in
a single unitary evolution, and checked the features of the asymmetric
cloner employed. In this subsection we consider how the quantum machine
achieves the goal of combining disentanglement and broadcasting in a united
way. The goal is achieved by copying both qubits separately.

Now both two qubits in the state $\left| \chi \right\rangle $ are cloned
according to the transformation defined by Eqs. (3) and (4) separately. Then
two copies $\rho _{a_{I}a_{II}}^{out}$ and $\rho _{b_{I}b_{II}}^{out}$ of
the entangled state $\left| \chi \right\rangle $ are produced. What we want
to do is to obtain a disentangling state and a copying state of the state $%
\left| \chi \right\rangle $. We check the inseparability of two copies $\rho
_{a_{I}a_{II}}^{out}$ and $\rho _{b_{I}b_{II}}^{out}$. The output density
matrices $\rho _{a_{I}a_{II}}^{out}$ and $\rho _{b_{I}b_{II}}^{out}$ are
given by 
\begin{equation}
\begin{array}{c}
\rho _{a_{I}a_{II}}^{out}=(\frac{(1+p^{2})^{2}}{N^{2}}\alpha ^{2}+\frac{q^{4}%
}{N^{2}}\beta ^{2})\left| 00\right\rangle \left\langle 00\right| +(\frac{%
(1+p^{2})^{2}}{N^{2}}\beta ^{2}+\frac{q^{4}}{N^{2}}\alpha ^{2})\left|
11\right\rangle \left\langle 11\right| \\ 
+\frac{(1+p^{2})q^{2}}{N^{2}}\left| 01\right\rangle \left\langle 01\right| +%
\frac{(1+p^{2})q^{2}}{N^{2}}\left| 10\right\rangle \left\langle 10\right| \\ 
+4\frac{p^{2}}{N^{2}}\alpha \beta \left| 00\right\rangle \left\langle
11\right| +4\frac{p^{2}}{N^{2}}\alpha \beta \left| 11\right\rangle
\left\langle 00\right|
\end{array}
\end{equation}
\begin{equation}
\begin{array}{c}
\rho _{b_{I}b_{II}}^{out}=(\frac{(1+q^{2})^{2}}{N^{2}}\alpha ^{2}+\frac{p^{4}%
}{N^{2}}\beta ^{2})\left| 00\right\rangle \left\langle 00\right| +(\frac{%
(1+q^{2})^{2}}{N^{2}}\beta ^{2}+\frac{p^{4}}{N^{2}}\alpha ^{2})\left|
11\right\rangle \left\langle 11\right| \\ 
+\frac{(1+q^{2})p^{2}}{N^{2}}\left| 01\right\rangle \left\langle 01\right| +%
\frac{(1+q^{2})p^{2}}{N^{2}}\left| 10\right\rangle \left\langle 10\right| \\ 
+4\frac{q^{2}}{N^{2}}\alpha \beta \left| 00\right\rangle \left\langle
11\right| +4\frac{q^{2}}{N^{2}}\alpha \beta \left| 11\right\rangle
\left\langle 00\right| .
\end{array}
\end{equation}
Again, it follows from Peres -Horodecki theorem that if $\frac{1-\sqrt{3}+%
\sqrt{2\sqrt{3}}}{2}\leq p\leq 1$, $\rho _{a_{I}a_{II}}^{out}$ is
inseparable for 
\begin{equation}
\begin{array}{c}
\frac{1}{2}-\sqrt{\frac{1}{4}-(\frac{(1+p^{2})(1-p)^{2}}{4p^{2}})^{2}}\leq
\alpha ^{2}\leq \frac{1}{2}+\sqrt{\frac{1}{4}-(\frac{(1+p^{2})(1-p)^{2}}{%
4p^{2}})^{2}},
\end{array}
\end{equation}
however $\rho _{b_{I}b_{II}}^{out}$ is separable for any values of $\alpha
^{2}$. So by choosing appropriate value of $p$ of the asymmetric cloner, the
copying state $\rho _{a_{I}a_{II}}^{out}$ and the disentangling state $\rho
_{b_{I}b_{II}}^{out}$ can be obtained in a single evolution.

For the copying state $\rho _{a_{I}a_{II}}^{out}$, the fidelity with respect
to the original entangled state $\left| \chi \right\rangle $ is examined.
The fidelity is defined as 
\begin{equation}
F=\left\langle \chi \right| \rho _{a_{I}a_{II}}^{out}\left| \chi
\right\rangle =\frac{(1+p^{2})^{2}}{N^{2}}-\frac{8pq^{2}}{N^{2}}\left|
\alpha \right| ^{2}\left| \beta \right| ^{2}.
\end{equation}
Obviously, the fidelity $F$ is dependent on the input entangled state $%
\left| \chi \right\rangle $. For the disentangling state $\rho
_{b_{I}b_{II}}^{out}$, the factors $s_{i}$ of qubits $b_{I}$ and $b_{II}$
are inspected. 
\begin{eqnarray}
\rho _{b_{I}}^{out} &=&Tr_{b_{II}}(\rho _{b_{I}b_{II}}^{out})  \nonumber \\
&=&\frac{2q}{N}(\alpha ^{2}\left| 0\right\rangle \left\langle 0\right|
+\beta ^{2}\left| 1\right\rangle \left\langle 1\right| )+\frac{p^{2}}{N}%
(\left| 0\right\rangle \left\langle 0\right| +\left| 1\right\rangle
\left\langle 1\right| )  \nonumber \\
&=&\frac{2q}{N}Tr_{a_{II}}(\left| \chi \right\rangle \left\langle \chi
\right| )+\frac{p^{2}}{N}I,
\end{eqnarray}
\begin{equation}
\rho _{b_{II}}^{out}=\frac{2q}{N}Tr_{a_{I}}(\left| \chi \right\rangle
\left\langle \chi \right| )+\frac{p^{2}}{N}I.
\end{equation}
It follows that $s_{b_{I}}=$ $s_{b_{II}}=\frac{2(1-p)}{N}$ (p=1-q). In the
range of $\frac{1-\sqrt{3}+\sqrt{2\sqrt{3}}}{2}\leq p\leq 1$, $s_{b_{I}}=$ $%
s_{b_{II}}\leq \frac{1}{\sqrt{3}}$. So the maximum value of closeness which
can be achieved by this process is $\frac{1}{\sqrt{3}}$ as in \cite{86,88}.
Of course, the maximum value $\frac{1}{3}$ of $s$ can be achieved in the
quantum machine by copying only one qubit.

Moreover it is worthwhile to notice that, the range of value of $p$ in the
case of copying two qubits separately is not simply equal to that in the
case of copying only one qubit as a result of the lost of quantum
information when the entanglement is broadcasted by local cloning.

Therefore by copying both two qubits asymmetrically it is possible to
combine disentanglement and broadcasting of entanglement in a single unitary
evolution. The fidelity of the output copying state with respect to the
input entangled state is state-dependent. While the scaling parameter $s$,
which can be achieved by the proposed quantum machine, has the same range as
in the work \cite{86,88}.

\section{Conclusion}

We have proposed a quantum machine, which for an input entangled state
produces a disentangling state and a copying state in a single unitary
evolution. The machine is based on the asymmetric cloner. The flow of
information in the cloning process is controlled by varying the parameter $p$
so that the quantum entanglement is partially retained in one copy of the
entangled state but erased in another. If using the 1$\rightarrow $2
asymmetric telecloning in the quantum machine, we can distantly send a
copying state of the entangled state to a receiver and a disentangling state
of it to another according to the requirement of information distribution.

To conclude, in this short note the disentanglement and broadcasting is
combined in a single evolution. We hope that it is helpful for understanding
entanglement and useful for further studying quantum information and quantum
computation.

\begin{center}
{\bf Acknowledgements}
\end{center}

This work has been financially supported by the National Natural Science
Foundation of China under the Grant No.10074072.


\begin{references}
\bibitem{bch}  C. H. Bennett, G. Brassard, S. Popescu, B. Schumacher, J. A.
Smolin, W.\ K. Wootters, Phys. Rev. Lett. 76, 722 (1996); D. Deutsch, A.
Ekert, R. Jozsa, C. Macchiavello, S. Popescu, and A. Sanpera, Phys. Rev.
Lett. 77, 2818(1996).

\bibitem{87}  V. Buzek, V.Vedral, M. B. Plenio, P. L. Knight, and M.
Hillery, Phys. Rev. A 55, 3327(1997).

\bibitem{85}  S. Bandyopadhyay and G. Kar, Phys. Rev. A 60, 3296(1999).

\bibitem{83}  D. R. Terno, Phys. Rev. A 59,3320(1999).

\bibitem{82}  T. Mor, Phys. Rev. Lett. 83,1451(1999).

\bibitem{1}  W. K. Wootters and W. H. Zurek, Nature (London) 299, 802(1982).

\bibitem{2}  A. K. Pati and S. L. Braunstein, Nature (London) 404, 164(2000).

\bibitem{3}  H. Barnum, C. M. Caves, C. A. Fuchs, R.Jozsa, and B.
Schumacher, Phys. Rev. Lett. 76, 2818 (1996).

\bibitem{88}  S. Ghosh, S. Bandyopadhyay, A. Roy, D. Sarkar, and G. Kar,
Phys. Rev. A 61,052301(2000).

\bibitem{7}  S. Bandyopadhyay, G. Kar, and A. Roy, Phys. Lett. A
258,205(1999).

\bibitem{8}  S. Ghosh, G. Kar, A. Roy, D. Sarkar, and U. Sen, Phys. Rev. A
64, 042114(2001).

\bibitem{84}  T. Mor and D. R. Terno, Phys. Rev. A 60,4341(1999).

\bibitem{86}  S. Bandyopadhyay, G. Kar, and A. Roy, Phys. Lett. A
258,205(1999).

\bibitem{83(3)}  D. Bruss, D. P. Divincenzo, A.Ekert, C. A. Fuchs, C.
Macchiavello, and J. A. Smolin, Phys. Rev. A 57, 2368 (1998).

\bibitem{99}  D. Bruss, Phys. Rev. A 60, 4344 (1999).

\bibitem{acta}  N. J. Cerf, Acta Phys. Slov. 48, 115(1998).

\bibitem{nj}  N. J. Cerf, Phys. Rev. Lett. 84, 4497 (2000).

\bibitem{12}  A. Peres, Phys. Rev. Lett. 77, 1413 (1996).

\bibitem{13}  M. Horodecki, P. Horodecki and R. Horodecki, Phys. Lett. A
223, 1 (1996).

\bibitem{177}  V. Buzek, M. Hillery and R. Bednik, Acta Phys. Slov. 48,
177(1998).
\end{references}
\end{document}